\documentclass[twocolumn,showpacs,preprintnumbers,prc,nofootinbib]{revtex4-2}
\usepackage{graphicx,color}
\usepackage{float}
\usepackage{bm}
\usepackage[pdftex,colorlinks=true, linkcolor = blue, citecolor=blue,urlcolor=blue, bookmarksnumbered=true, bookmarksopen=true]{hyperref}
\begin{document}

\title{Time Evolution of Prompt Gamma-Ray Emission in $^{252}$Cf(sf) and $^{233,235}$U($n$,f) Reactions}

\author{$^{1,2}$Patrick Talou}
\author{$^1$Amy~E. Lovell}
\author{$^1$Ionel Stetcu}
\author{$^1$Gencho Rusev}
\author{$^1$Toshihiko Kawano}
\author{$^3$Marian Jandel}
\affiliation{$^1$Los Alamos National Laboratory, Los Alamos, NM 87545, USA}
\affiliation{$^2$Stardust Science Labs, Santa Fe, NM 87507, USA}
\affiliation{$^3$University of Massachusetts Lowell, Lowell, Massachusetts 01854, USA}

\newcommand{\CGMF}{$\mathtt{CGMF}$}
\newcommand{\CGMFtk}{$\mathtt{CGMFtk}$}
\newcommand{\GEANT}{$\mathtt{GEANT4}$}

\newcommand{\COH}{$\mathtt{CoH}$}
\newcommand{\TALYS}{$\mathtt{TALYS}$}
\newcommand{\EMPIRE}{$\mathtt{EMPIRE}$}

\newcommand{\g}{$\gamma$}
\newcommand{\gray}{$\gamma$-ray}
\newcommand{\grays}{$\gamma$ rays}
\newcommand{\aveNg}{$\overline{N}_\gamma$}

\newcommand{\etal}{{\it et al.}}

\date{\today}

\begin{abstract}
We investigate the time evolution of prompt fission \g\ emission due to the presence of ns to ms isomers in the fragments produced in the neutron-induced fission of $^{233,235}$U and in the spontaneous fission of $^{252}$Cf. Calculations performed with the \CGMF\ fission event generator are compared with recent experimental data on $^{252}$Cf(sf) and $^{235}$U($n$,f) obtained with the DANCE+NEUANCE setup at the Los Alamos Neutron Science Center. Of particular interest are the average \gray\ energy spectrum as a function of time since scission, $\phi(\epsilon_\gamma,t)$, the increase of the average \gray\ multiplicity over time, \aveNg $(t)$, and its evolution in time as a function of the fission fragment mass, \aveNg(A,$t$). The time evolution of isomeric ratios in post-neutron emission fission fragments can be defined and used to test and reveal some deficiencies in our knowledge of the low-lying levels of neutron-rich nuclei produced in fission reactions. 
\end{abstract}
\maketitle

\section{Introduction} \label{sec:introduction}

Nuclear fission events are typically accompanied by the emission of prompt neutrons and \grays\ within a few ns of the separation or scission of the fragments. Although those prompt \grays\ are certainly of lesser importance than their prompt neutron counterpart in supporting the fission chain reaction, they are nevertheless relevant for a variety of applications, including local \g\ heating in reactors~\cite{Blaise:2014} or active interrogation of special nuclear materials~\cite{Finch:2021}. They are also highly useful and informative signatures of the fission process~\cite{Wilson:2021}. When analyzed in correlation with the prompt neutron and fission fragment characteristics, they can provide tight constraints on specific quantities such as the initial spin distribution of the fission fragments~\cite{Stetcu:2014,Vogt:2021,Stetcu:2021,Bulgac2021rel}.

Late Prompt Fission Gamma rays  (LPFGs) are emitted typically within a few ns to a few ms from the isomeric states of fragments populated after prompt neutron emission in the fission process. Those are not to be confused with $\beta$-delayed \grays\ that are emitted only after the fission products have further decayed through a $\beta^-$ process, which typically takes longer than a ms. The study of these LPFGs is important for the correct interpretation of nuclear diagnostic subcritical experiments~\cite{Gomez:2019,Talou:2016} or the inference of fission product yields from \g\ coincidence measurements~\cite{Wilson:2017,Fotiades:2019}.

In the present paper, we study LPFGs in the case of $^{252}$Cf spontaneous fission and in the neutron-induced fission of $^{233,235}$U. Unique and invaluable experimental data on the prompt fission \gray\ spectrum as a function of time have been obtained~\cite{Rusev:2020,Rusev:2025} using the DANCE+NEUANCE setup at the Los Alamos Neutron Science Center (LANSCE) facility at Los Alamos. Direct comparison of the measured data with theoretical predictions is made difficult by the complexity of the experimental apparatus. Instead, the forward propagation of model predictions through a detailed \GEANT\ simulation of the experimental setup is performed and propagated simulated events compared to the data. The model calculations are performed using the \CGMF\ Monte Carlo code that follows the decay of fission fragments on an event-by-event basis, providing all required information on prompt \grays, including their time of emission.

We review our theoretical approach and provide a description of our \CGMF\ simulation code in Sec.~\ref{sec:theory}. We also briefly include some details about the \GEANT\ model developed and used to simulate the response of the DANCE + NEUANCE detector setup. In Sec.~\ref{sec:results}, we present and discuss our results in light of their impact on our knowledge of the nuclear structure of the neutron-rich nuclei produced in the fission reactions studied here. 

\section{Theory \& Modeling} \label{sec:theory}

\subsection{Theoretical Description} \label{sec:theoreticalDescription}

Here we provide a brief description of our fission fragment decay model with emphasis on the parts relevant to the present discussion on prompt \g\ emission. The \CGMF\ code~\cite{Talou:2021} has been used extensively to study various aspects of the fission process: multi-chance fission~\cite{Lovell:2021}, fission product yields~\cite{Okumura:2018}, prompt fission \grays~\cite{Stetcu:2014}, correlated anisotropies between fission fragments and prompt neutrons~\cite{Lovell:2020}, and neutron-\g\ correlations~\cite{Marin:2021}. An overview of correlated fission studies can also be found in Ref.~\cite{Talou:2018}.

The nuclear fission process can produce hundreds of scission fragments characterized by their mass, $A$, and charge, $Z$, numbers. They are formed in a wide range of configurations in terms of kinetic energy, KE, excitation energy, $U$, spin, $J$, and parity, $\pi$. From those initial conditions right after scission, the fragments typically emit neutrons and photons to shed their intrinsic and rotational excitation energies until they reach a long-lived state. Most neutron and photon emissions occur within a very short amount of time after scission (less than $10^{-14}$ sec). This picture is somewhat complicated by the presence of isomeric states in post-neutron emission fragments that delay the emission of photons according to their half-lives. 

The prompt fission \gray\ spectrum at a given time $t$ can be written as a sum over all fragments produced in the fission process and their $\gamma$ decay chains
\begin{eqnarray} \label{eq:PFGS}
\phi(\epsilon_\gamma,t) = \sum_i{Y_i(A,Z) \sum_{\nu}{P_i(\nu)} \phi_i(A-\nu,Z,\epsilon_\gamma,t)},
\end{eqnarray}
where $Y_i(A,Z)$ are the initial distributions of the scission fragments in mass and charge, $P_i(\nu)$ is the neutron multiplicity distribution for this fragmentation, and $\phi_i$ represents the time-dependent $\gamma$ emission spectrum for the fragmentation $(A-\nu,Z)$. This seemingly simple equation hides many complexities and details. The fragments $(A,Z)$ are produced in a range of initial conditions, hence
\begin{eqnarray}
Y_i(A,Z) \equiv Y_i(A,Z,{\rm KE},U,J,\pi),
\end{eqnarray}
in which the dependences of the initial fragment yields on mass, $A$, charge, $Z$, kinetic energy, KE, initial excitation energy, $U$, initial angular momentum, $J$, and parity, $\pi$, are given explicitly. Each fragment can then emit one or more neutron(s) according to a probability $P_i(\nu)$, leading to a residual, post prompt-neutron emission fragment with a residual $(U,J,\pi)'$ distribution, where $U'=U-\epsilon_n-S_n$ for a neutron emitted with energy $\epsilon_n$. Once the probability of neutron emission is exhausted, a photon cascade would typically conclude this decay chain until the ground state of the residual fragment is reached. If all photons were to be emitted practically in coincidence with the scission event, then we could drop the time $t$ from Eq.~(\ref{eq:PFGS}). This is not the case, however, as the decay cascade can also populate long-lived isomers whose half-lives determine the time evolution of further \g\ emissions.

\subsection{The \CGMF\ code} \label{sec:CGMF}

The \CGMF\ code~\cite{Talou:2021} was developed to specifically follow in detail the decay of fission fragments through the emission of neutrons and photons, and is therefore particularly suited to compute the time-dependent $\gamma$ spectrum as expressed in Eq.~(\ref{eq:PFGS}). \CGMF\ is a Monte Carlo implementation of the statistical nuclear reaction theory as expressed through the Hauser-Feshbach formalism, sampling initial scission fragment conditions and the neutron and photon emission probabilities at each stage of the decay. It relies on several assumptions and input parameters that are briefly reviewed here, especially those which could impact our present results and conclusions. More details can be found in Ref.~\cite{Talou:2021} and references therein.

The initial scission yields are described as a weighted sum of 5 Gaussian distributions that can be understood as fission modes often used to describe experimental fragment yields. This representation is simple yet powerful enough to be used successfully in the present work. An effort to evaluate those initial fragment yields in a more accurate fashion is underway~\cite{Lovell:2021}. The present calculations are similar enough to the final evaluated results and would not impact the conclusions presented in this paper. 

\begin{figure}
    \centering
    \includegraphics[width=\columnwidth]{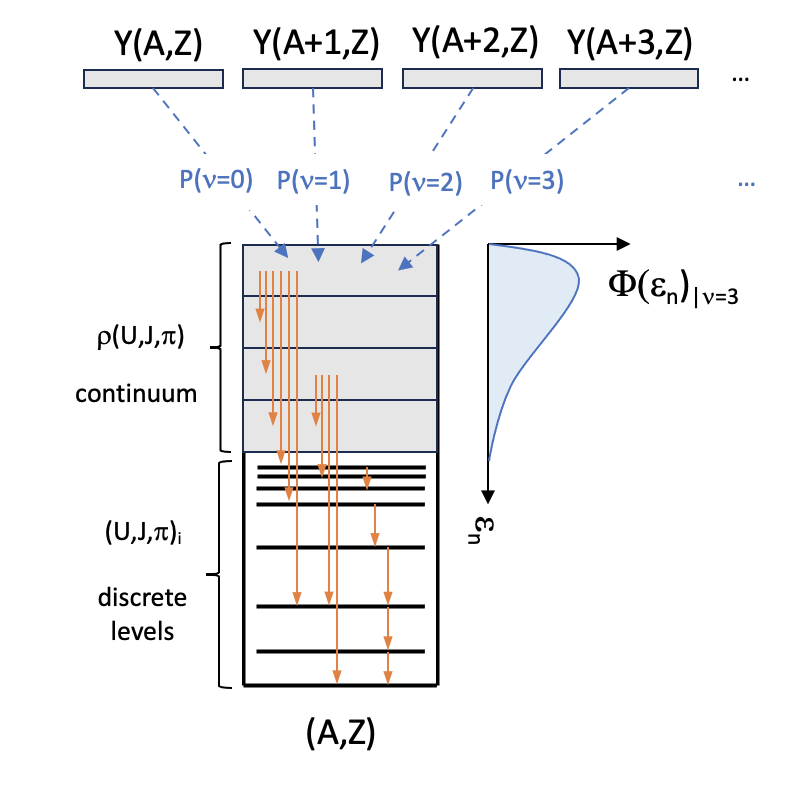}
    \caption{A schematic view of the decay chains followed by \CGMF\ to infer all possible neutron and \g\ emissions from scission fragments initially produced in a fission reaction.}
    \label{fig:decay_chain}
\end{figure}

In \CGMF, the neutron-gamma emission competition is handled through the Hauser-Feshbach statistical equations and emission probabilities. In the case of the decay of fragments produced in low-energy fission reactions, the initial excitation energy in the fragments rarely ever exceeds about 30 MeV. At this energy, only neutron and \g\ emissions are allowed, the Coulomb barrier hindering the evaporation of charged particles such as protons and $\alpha$ particles. Note that we are only discussing the evaporation of charged particles and not their possible emission through a dynamical process as observed in ternary fission, for instance. Therefore, the probability of emitting a neutron or a \gray\ is given by
\begin{eqnarray}
P_{n,\gamma} = \frac{\Gamma_{n,\gamma}}{\Gamma_n+\Gamma_\gamma},
\end{eqnarray}
where $\Gamma_{n,\gamma}$ are the decay widths for the neutron and \g\ emission, respectively. The neutron widths are calculated from the $S$-matrix obtained with an optical model calculation. Default \CGMF\ runs use the global spherical optical model of Koning-Delaroche~\cite{Koning:2003}. Prompt fission neutron spectra calculated with \CGMF\ in this default mode tend to be softer than reported experimentally. While this remains an unresolved issue at this point, it has a negligible impact on the study of prompt \grays. More importantly than the neutron energy spectrum, the neutron probability distribution, P($\nu$), that indicates how many neutrons are emitted from the initial fragments, is an important quantity that is well reproduced in the present calculations.

The \g\  emission widths, $\Gamma_\gamma$, or probabilities, $P_\gamma$, are calculated using a strength function formalism at higher excitation energies, where the nucleus levels can only be described using a continuum representation characterized by the density $\rho(U,J,\pi)$. The parity distribution is assumed to be equiprobable between positive and negative values. The spin distribution is given as follows

\begin{eqnarray} \label{eq:spin}
P(J) =\frac{1}{C}(2J+1)\exp{\left\{\frac{J(J+1)}{2B^2(A,Z,T)}\right\}},
\end{eqnarray}
where $C$ is a normalization coefficient, $T$ is the temperature in the fragment $(A,Z)$, and $B$ is the temperature-dependent spin cut-off parameter given as
\begin{eqnarray} \label{eq:spincutoff}
B^2(A,Z,T) = \alpha \frac{\mathcal{I}_0(A,Z)T}{\hbar^2}.
\end{eqnarray}
$\mathcal{I}_0$ is the moment of inertia of the fission fragment, and $\alpha$ is an adjustable, global, energy-dependent parameter that is tuned to reproduce average prompt \gray\ properties such as the the average \gray\ multiplicity and isomeric ratios in fission products following $\beta$ decay. 

At lower excitation energies, \CGMF\ uses the spectroscopic knowledge of the nuclear structure of the populated fragments as reported by the nuclear structure evaluation community in the ENSDF database~\cite{ENSDF}. This database contains the experimentally-known discrete levels for all isotopes of the nuclear chart. An important caveat is that \CGMF\ does not use ENSDF directly but instead uses the RIPL-3 database~\cite{RIPL3} of discrete levels. It is a curated version of ENSDF that complements it to provide a reasonable match with the continuum level density used in nuclear reaction codes such as \COH, \TALYS, \EMPIRE\ or even \CGMF. This important question will be revisited in Sec.~\ref{sec:results} when the importance of missing or erroneous discrete levels to reproduce the experimental spectrum observed in DANCE will be addressed.

Figure~\ref{fig:decay_chain} provides a schematic view of the decay chain process as followed by \CGMF. First, initial scission fragment yields, $Y(A,Z)$, are sampled as a function of their mass $A$ and charge $Z$. In addition, they are produced in a wide range of kinetic energy, excitation energy, spin, and parity values, which is not indicated in this figure for sake of clarity. In this sequence, neutrons are mostly emitted prior to \grays. In order to study the \g\ spectrum emitted from one particular fragment $(A,Z)$, as in Fig.~\ref{fig:decay_chain}, we first need to follow the neutron emission probabilities, P$(\nu)$, from all initial scission fragments that feed the post-neutron fragment of interest. In addition, this feeding happens for a range of residual excitation energies and spins, which can impact the final observed prompt \g\ spectrum stemming from this particular fragment. In addition, this fragment can itself decay by emitting one or more neutrons, taking some flux away from its own \g\ decay channel.

A typical run of \CGMF\ records millions of fission events characterized by the mass and charge numbers of the complementary fragments, the number (or multiplicity), the energy and the momentum vectors of the neutrons emitted from each fragment (and of any possible pre-scission neutron emission), and similar data for the emitted \grays. For this particular study, the time of emission of each \g\ ray is also recorded through a special user input option. The result is a large ASCII file that is then analyzed using the \CGMFtk\ suite of python classes and routines~\cite{Talou:2021}, as well as any extra python routines developed for the present work. Those simulated data are recorded on an event-by-event basis, hence provide a powerful tool to slice and project this large multi-dimensional dataset in various ways to study any type of correlations. It is particularly useful in the present context when we want to trace back the origin of particular \g\ lines observed (or not) in the DANCE+NEUANCE experiments.

Among the discrete levels known at lower excitation energies, as represented in Fig.~\ref{fig:decay_chain}, some of them are isomers with a half-life of a few ns to a few seconds or more. Their feeding and further decay is recorded in \CGMF. Their time of emission in a \CGMF\ event is randomly sampled according to their reported half-life. 

\subsection{Detector Response Simulation} \label{sec:DANCE}

\label{sec:detectorResponse}

Here we briefly describe the experimental setup used to measure and analyze late prompt fission \grays. A more complete and in-depth description can be found in Rusev \etal~\cite{Rusev:2020,Rusev:2025}.

The Detector for Advanced Neutron Capture Experiments (DANCE)~\cite{Ullmann:2003} is a $4\pi$ high-granularity, high-efficiency \gray\ calorimeter comprised of 160 BaF$_2$ detectors, installed at the Los Alamos Neutron Science Center (LANSCE). It was used to measure the prompt fission \grays\ in the $^{252}$Cf spontaneous fission and $^{235}$U($n$,f) reactions. The $^{235}$U target was 26 mg/cm$^2$ thick (with neutron beam) and the $^{252}$Cf source had an $\alpha$ activity of 0.5 $\mu$Ci (without a beam). The neutron beam at LANSCE is produced via spallation of 800-MeV protons impinging on a thick tungsten target, and collimated to create a 5-mm (FWHM) beam at the center of DANCE, 20.25m from the water moderator placed near the tungsten target.

For these particular experiments, the default configuration of DANCE with a $^6$LiH shell installed in the central cavity was replaced with the NEUtron detector for DANCE (NEUANCE) to identify fission events~\cite{Jandel:2018}. NEUANCE comprises 21 stilbene detectors arranged cylindrically around the target, which can detect neutrons with energy greater than 0.5 MeV.

Both DANCE and NEUANCE detector arrays have a fast time response and can be used to detect and study coincidence events within a 5~ns time window. Fission events are recorded when a neutron triggers a NEUANCE detector in coincidence with \grays\ firing at least three DANCE detectors with a total \gray\ energy greater than 3 MeV. Various types of background have to be removed to clearly identify the detected \grays\ as prompt fission particles. In particular, a multiplicity cut is applied to eliminate events due to cosmic rays, and an energy cut on the total \gray\ energy, $E_{\gamma}^{total}$, to remove capture events of the barium isotopes that produce \grays\ with a total \g\ energy from 4.72 to 9.11 MeV. The remaining background can be cleaned up using a method~\cite{ODonnell:2016} originally developed for prompt fission neutron measurements carried out with the Chi-Nu experimental setup at LANSCE.

Another filter is applied to eliminate events that see two or three neighboring DANCE detectors firing simultaneously, most likely due to Compton scattering of a \gray\ from one detector to another, or pair creation and subsequent positron annihilation. 

\begin{figure}
\centering
\includegraphics[width=\columnwidth]{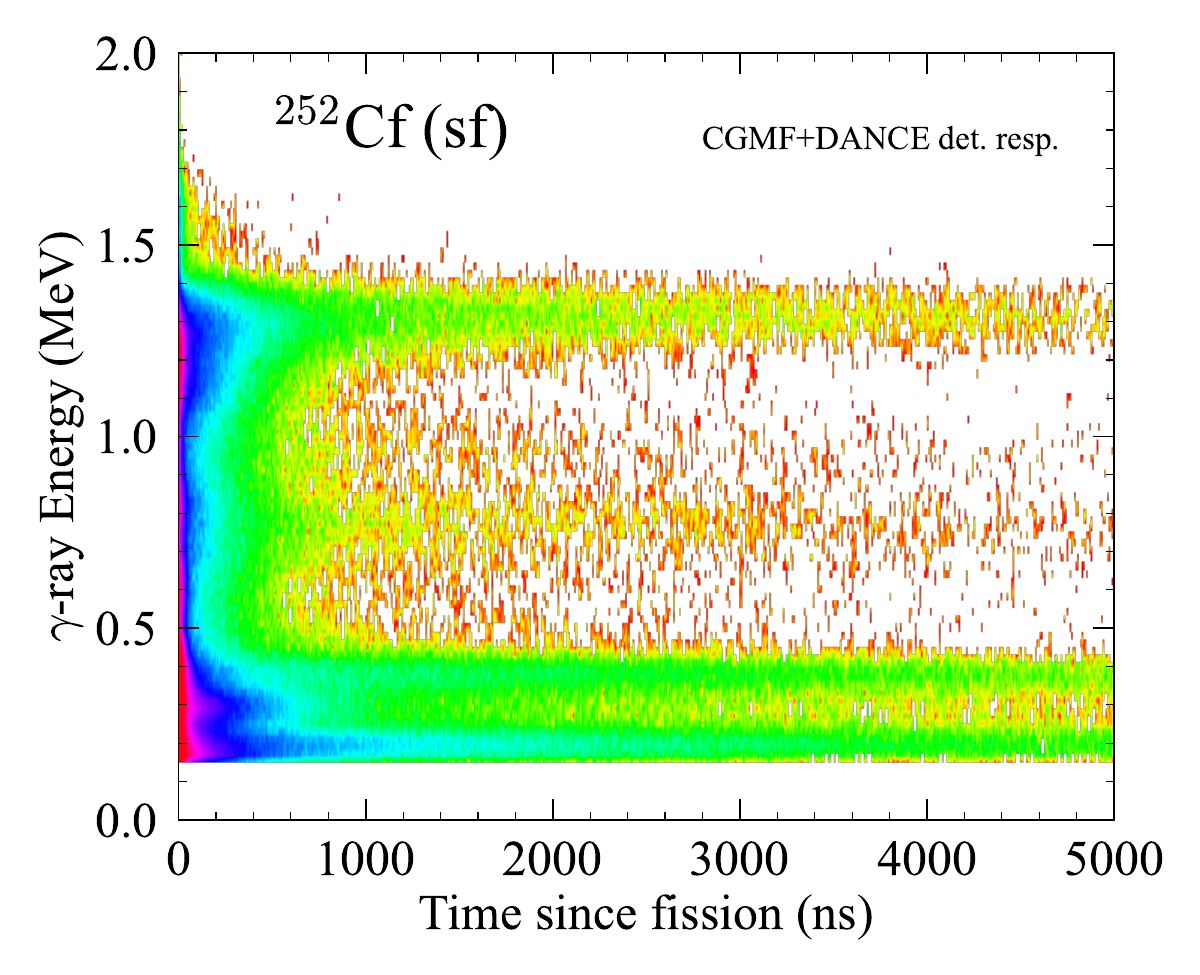}
\caption{Prompt \gray\ spectrum of $^{252}$Cf(sf) as a function of time since fission, as predicted by \CGMF\ and processed through the DANCE+NEUANCE detector response simulated in \GEANT.}
\label{fig:2Dspec_CGMF_DANCE}
\end{figure}

Models of both the DANCE and NEUANCE detector arrays have been implemented in the \GEANT\ transport code~\cite{Agostinelli:2003} to simulate a realistic detector response. The results of \CGMF\ calculations in the form of large Monte Carlo history files can then be used as input in \GEANT\ and processed model calculations compared to DANCE experimental data. An example of the time-dependent \gray\ energy spectrum simulated by \CGMF\ and processed through \GEANT\ is shown in Fig.~\ref{fig:2Dspec_CGMF_DANCE} in the case of $^{252}$Cf(sf).

\section{Results} \label{sec:results}

We now turn to the results obtained for the neutron-induced fission reactions on $^{233,235}$U and the spontaneous fission of $^{252}$Cf, and compare available DANCE+NEUANCE experimental data with the \CGMF\ simulated results propagated through our \GEANT\ simulation of the experimental setup. Note that by default, \CGMF\ uses the input data file labelled ``RIPL-3" (circa 2015) to read in the evaluated nuclear structure of all fission fragments. We will discuss the impact of using different nuclear structure input files, e.g., ENSDF (2024) and RIPL-3 (circa 2020), on some of our results.

\subsection{Time-dependent Gamma-ray Spectrum} \label{sec:gspec}

\begin{figure}
    \centering
    \includegraphics[width=\columnwidth]{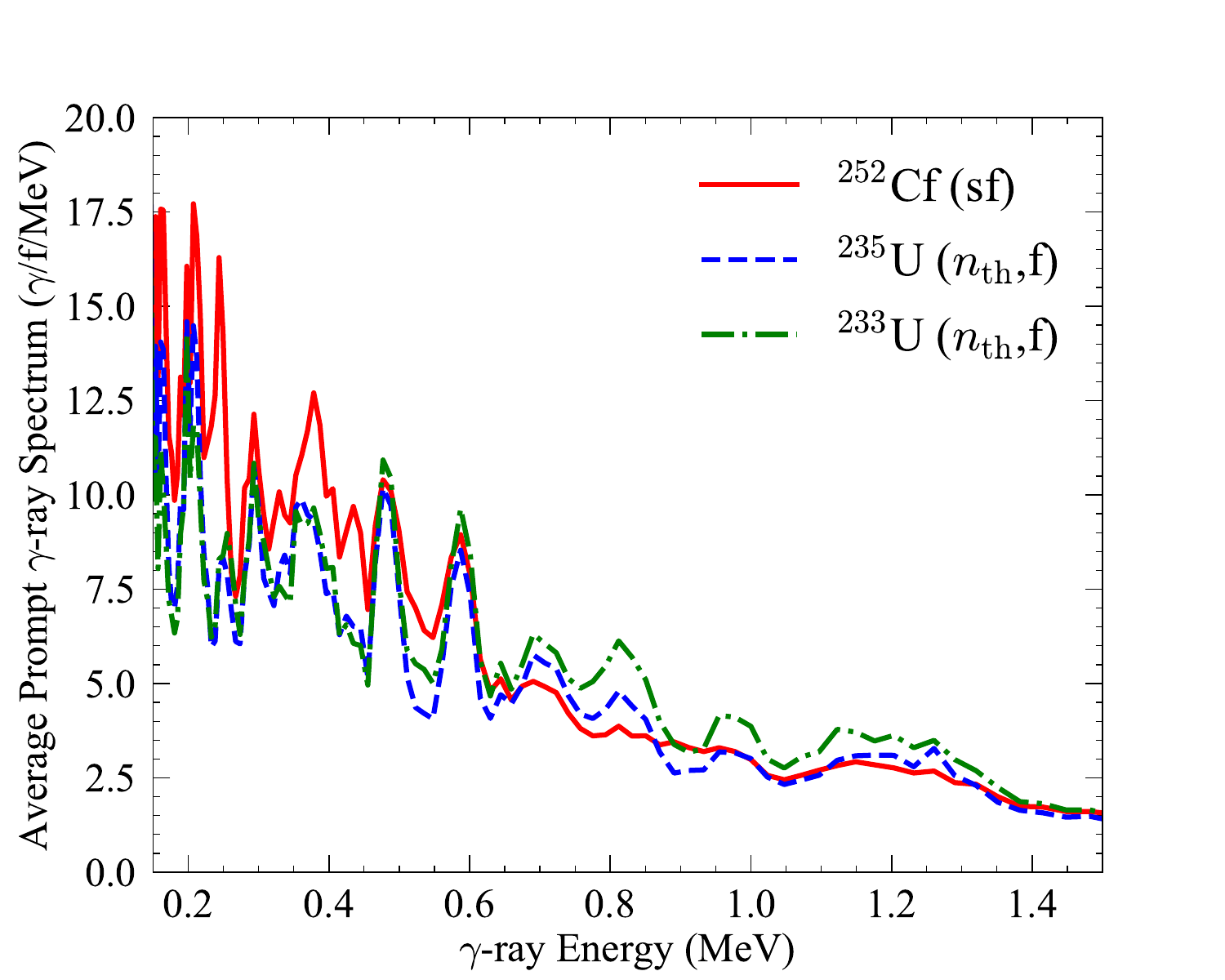}
    \caption{The average prompt fission \gray\ spectra calculated with \CGMF\ for the three fission reactions studied in this work: spontaneous fission of $^{252}$Cf and thermal-neutron-induced fission on $^{233}$U and $^{235}$U.}
    \label{fig:pfgs_alltimes}
\end{figure}

The low-energy part of the average prompt fission \gray\ spectrum integrated over time is shown in Fig.~\ref{fig:pfgs_alltimes} for all three fission reactions studied. The figure is cut at 150 keV, which is a typical energy threshold for many fission \gray\ measurements, including the ones performed at DANCE. The strong fluctuations observed in this low-energy part of the energy spectrum are characteristics of \g\ transitions between low-lying excited states in dominant fission fragments, and have been clearly identified experimentally in several fission reactions~\cite{Billnert:2013,Oberstedt:2013,Oberstedt:2015,Gatera:2017}. Many of these peaks are common to $^{252}$Cf(sf) and the neutron-induced fission reactions on $^{233,235}$U, demonstrating a strong overlap between fission fragments produced in those reactions. Significant differences can be seen though, even between the two uranium isotopes, with more \grays\ emitted from $^{233}$U than $^{235}$U in the 0.7 to 1.5 MeV region. 

\begin{figure}
    \centering
    \includegraphics[width=\columnwidth]{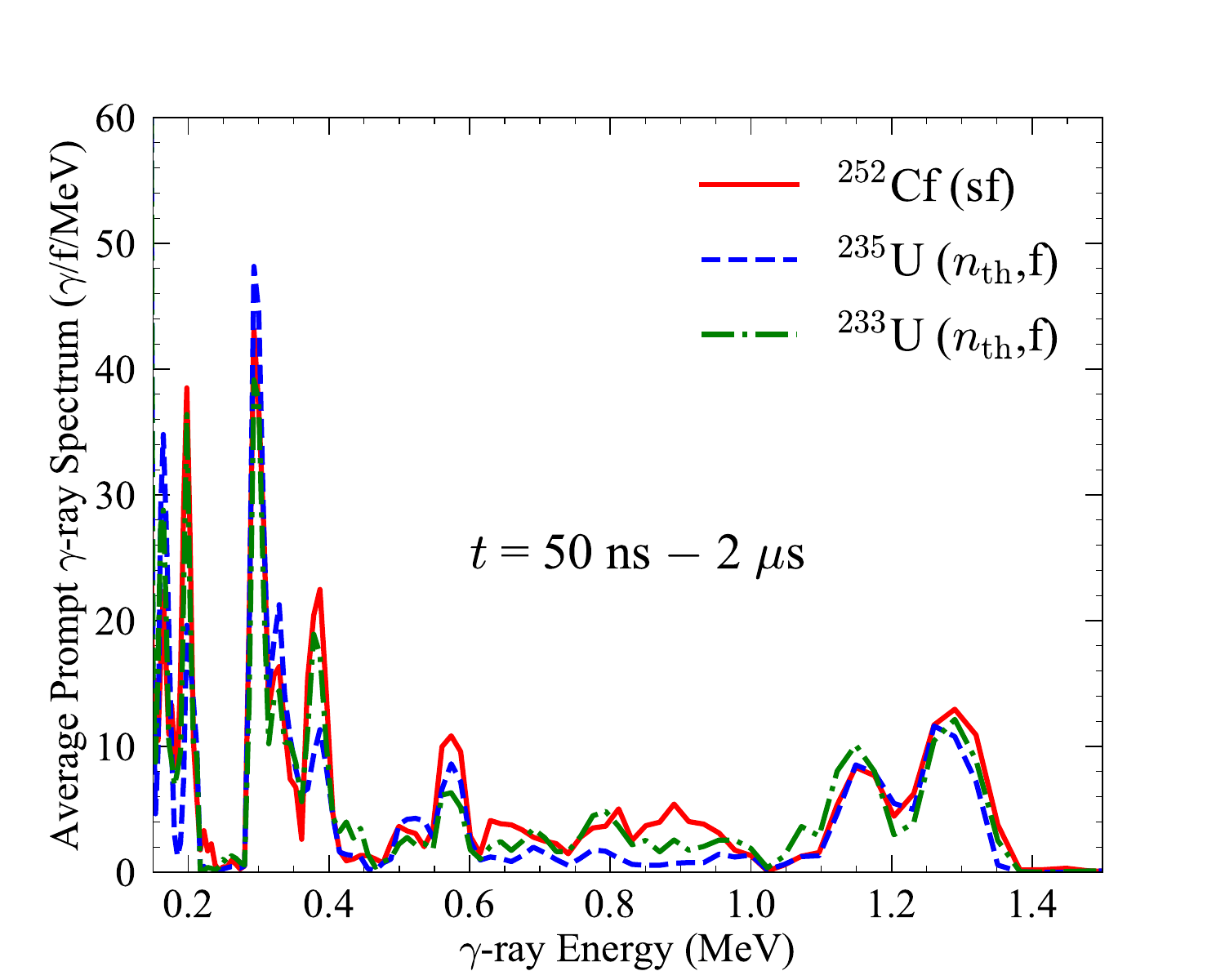}
    \caption{The same \CGMF-calculated average prompt fission \gray\ spectra as in Fig.~\ref{fig:pfgs_alltimes} but now considering only emission times between 50~ns and 2~$\mu$s.}
    \label{fig:pfgs_50ns-2us}
\end{figure}

Let us now look at what happens past the first burst of prompt fission \g\ emission and study the same spectra for \grays\ emitted between 50~ns and 2~$\mu$s after the fission event, as shown in Fig.~\ref{fig:pfgs_50ns-2us}. Strong features appear in all three spectra, corresponding to particular \g\ transitions from isomeric states with half-lives roughly corresponding to this time range. Some prominent peaks appear for new \gray\ energies. A rapid search for the isotopes that contribute to this spectrum reveals the significant role of ns isomers in Te isotopes with masses ranging from 130 to 135. These isomers will be discussed in more detail below.

\begin{figure}
    \centering
    \includegraphics[width=\columnwidth]{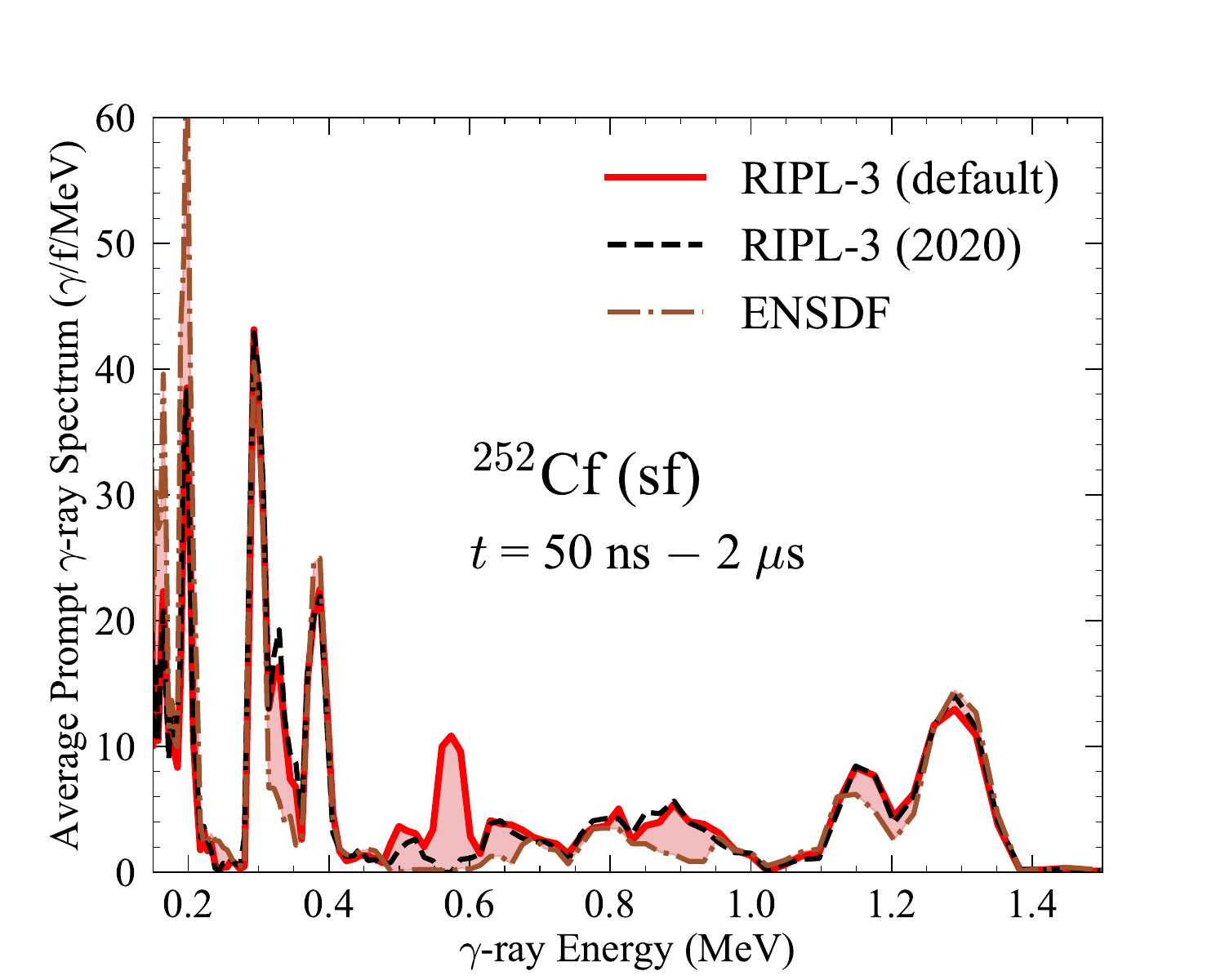}
    \caption{Same as Fig.~\ref{fig:pfgs_50ns-2us} but where we study the impact of the nuclear structure input file to the \CGMF-calculated average prompt fission \gray\ spectrum. The red-colored area denotes the difference between the default CGMF calculation using the RIPL-3 (2015) structure file and the one using the ENSDF file.}
    \label{fig:pfgs_50ns-2us_struct}
\end{figure}

As indicated in Fig.~\ref{fig:decay_chain}, nuclear level densities at high excitation energies in the fission fragments give way to a set of discrete levels at lower excitation energies. Those states have been identified up to a certain maximum excitation energy, which is different for all nuclei. Our knowledge of the spin and parity assignments of those levels also vary from nucleus to nucleus. As mentioned earlier, the nuclear structure evaluations present in ENSDF are often complemented or modified before they can be used in nuclear reaction codes like \CGMF, to ensure a smooth transition between the discrete and continuum representations of the nuclear levels and completeness of $(J,\pi)$ level assignments required by the code. By default, \CGMF\ uses the 2015 version of the RIPL-3 discrete levels database~\cite{RIPL3}, and imports the low-lying levels, their excitation energies, spins, parities and half-lives, reported for each nucleus in the nuclear chart. Obviously, not every nucleus is known with the same accuracy or level of confidence. In Fig.~\ref{fig:pfgs_50ns-2us_struct}, we show the same time-gated \gray\ spectrum as in Fig.~\ref{fig:pfgs_50ns-2us}-- for $^{252}$Cf(sf) only-- but obtained using different nuclear structure input files. Significant differences appear between the three spectra represented, in particular near 0.55 MeV where a distinct and strong peak appears in the case of RIPL-3, but neither in RIPL-3 (2020) nor in ENSDF. Other differences appear between both RIPL-3 results and the ENSDF one, in particular near 0.35, 0.9 and 1.15 MeV.

\begin{figure}[hb]
    \centering
    \includegraphics[width=\columnwidth]{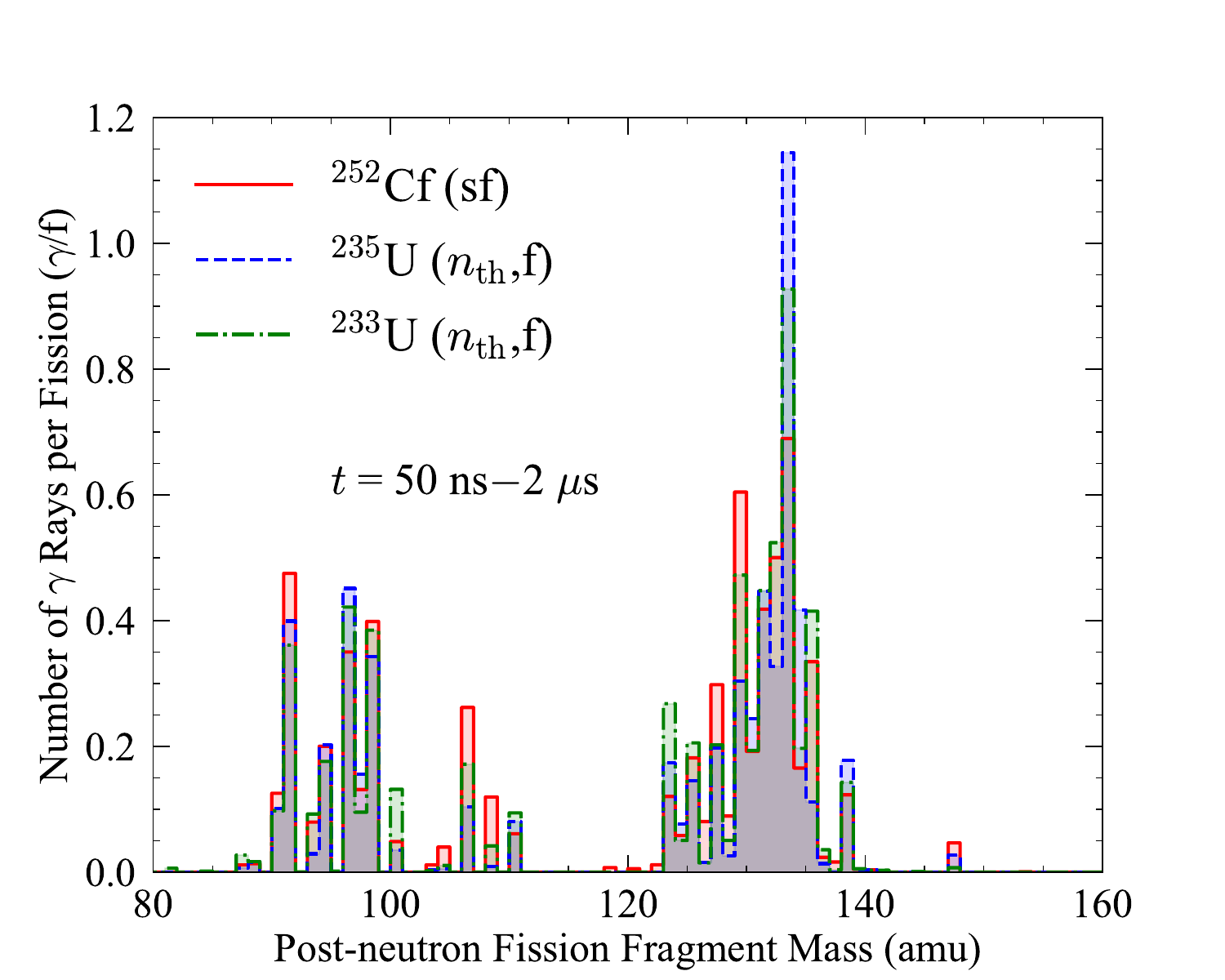}
    \caption{Calculated number of prompt fission \grays\ emitted per fragment within 50~ns and 2~$\mu$s as a function of the mass of the emitting post-neutron fission fragment. The default RIPL-3 discrete level files were used in these calculations.}
    \label{fig:NgAf_50ns-2us}
\end{figure}

One way to investigate those differences is to identify which isotopes among the hundreds of fission fragments produced contribute the most to this time-gated spectrum. The result is shown in Fig.~\ref{fig:NgAf_50ns-2us} where the number of \grays\ emitted in the 50~ns to 2~$\mu$s time window after fission is shown as a function of the mass of the emitting fission fragment. 

\begin{figure}
    \centering
    \includegraphics[width=\columnwidth]{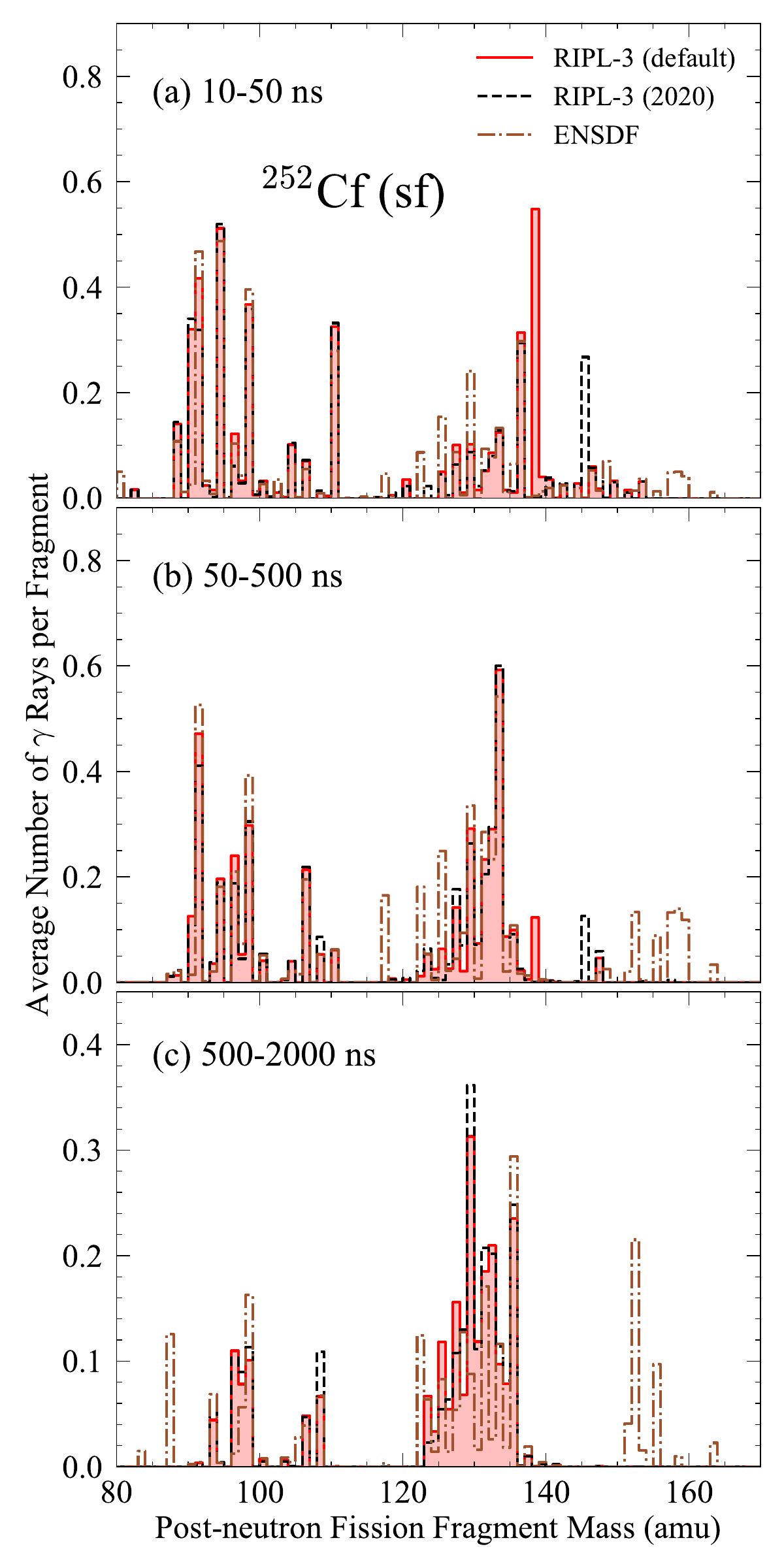}
    \caption{Number of \grays\ emitted per post-neutron fission fragment calculated with \CGMF\ for three time intervals: (a) 10-50 ns, (b) 50-500 ns, and (c) 500-2000 ns. These results are to be compared with the experimental results shown in Fig.~13 of John \etal~\cite{John:1970}.}
    \label{fig:NgAt-threeTimeIntervals}
\end{figure}

Of particular interest is the dependence of these predicted \g\ spectra on the choice of the nuclear structure file used in \CGMF. Figure~\ref{fig:NgAt-threeTimeIntervals} shows such a dependence in the case of $^{252}$Cf(sf) for the three nuclear structure files considered in this work. Three time intervals were considered: (a) 10-50 ns, (b) 50-500 ns, and (c) 500-2000 ns, to more easily compare with the experimental data in Fig.~13 of John \etal~\cite{John:1970}. While most peaks are more or less present in all calculated cases, albeit with different intensities, a specific group of \g\ peaks in $A=150-160$ fragment mass region only appears when using the ENSDF nuclear structure file.Those peaks are due mostly to isomeric decays identified in $^{153,156}$Nd and $^{159,160}$Sm isotopes. They are also present in the experimental data of John \etal~\cite{John:1970}, while some calculated peaks, e.g., near mass 85, are missing in the experiment. The experiment by John \etal\ is expected to be missing some strength due to the upper limit of the time coincidence window that would cut some longer-lived states from contributing to the signal. 

\begin{figure}
    \centering
    \includegraphics[width=\columnwidth]{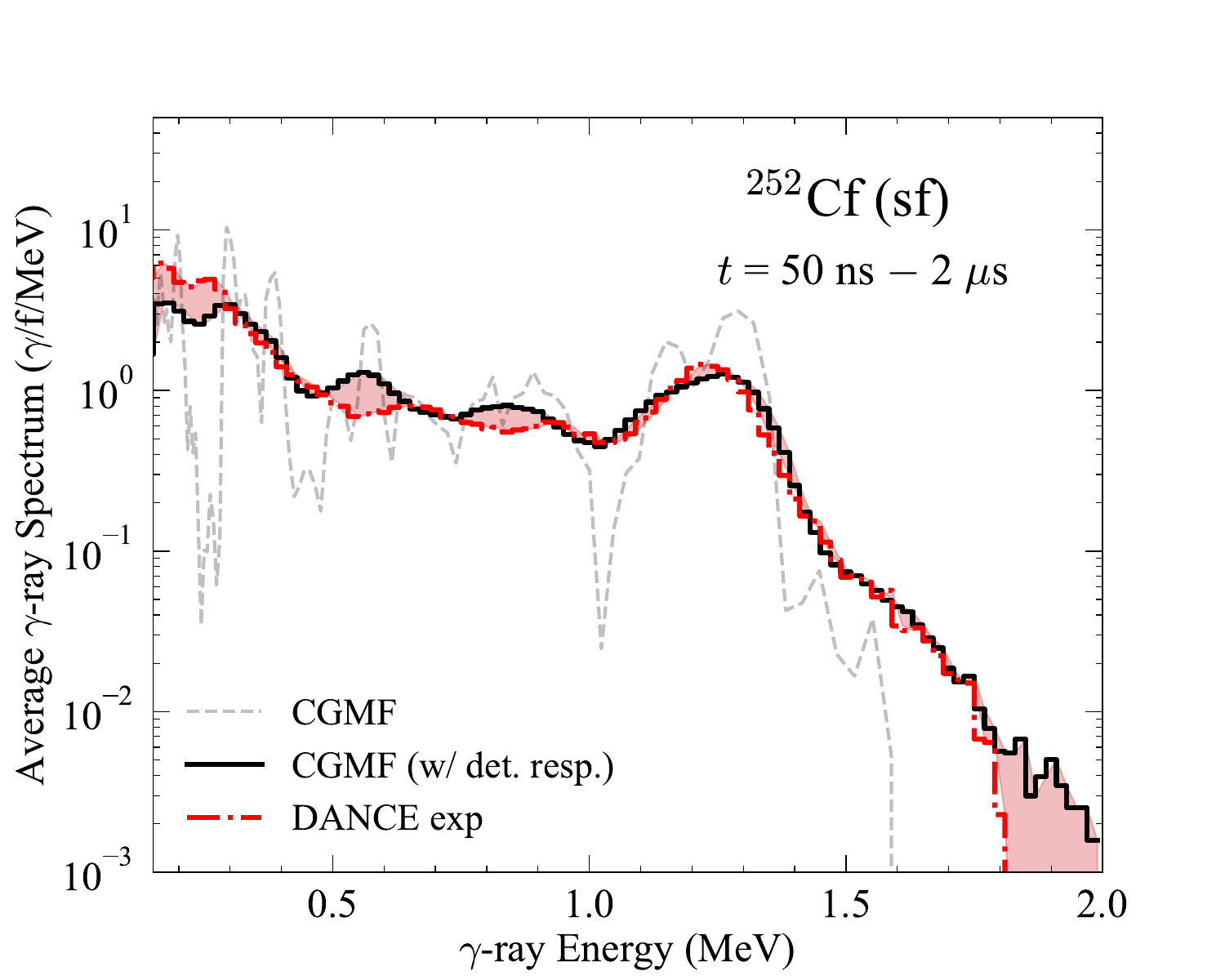}
    \caption{The DANCE experimental spectrum (red) is compared to the \CGMF-calculated spectrum (black) processed through the \GEANT\ simulation of the DANCE detector response. The ``raw" \CGMF\ spectrum is also shown (light dashed grey).}
    \label{fig:pfgs_detresp}
\end{figure}

\CGMF\ event-by-event results provide all necessary data to correlate each predicted peak in the \g\ spectrum to specific transitions in the underlying nuclear structure of the fission fragments. Although our analysis could be pushed further to determine the exact origin of those peaks, it is time to turn toward experimental data to validate or invalidate those simulated results. As discussed in Sec.~\ref{sec:detectorResponse}, the DANCE+NEUANCE detector setup was able to measure the time dependence of the \gray\ spectra for the three reactions considered here\footnote{The experimental DANCE data for $^{233}$U($n$,f) are still preliminary and will not be discussed here.}. However, DANCE does not have the energy resolution necessary to disentangle all peaks predicted by \CGMF. This is visible in Fig.~\ref{fig:pfgs_detresp} in the case of $^{252}$Cf spontaneous fission.

Several important observations can be made on the results shown in Fig.~\ref{fig:pfgs_detresp}. First, the \GEANT\ simulation of the DANCE+NEUANCE detector response forward propagating the \CGMF-calculated \grays\ smoothes out most of the fluctuations observed in Fig.~\ref{fig:pfgs_50ns-2us}, also shown in dashed gray in Fig.~\ref{fig:pfgs_detresp}. However, the overall shape of the spectrum is very well reproduced, validating the bulk of the \CGMF\ calculations of this time-gated \gray\ spectrum. 

\begin{figure}
    \centering
    \includegraphics[width=\columnwidth]{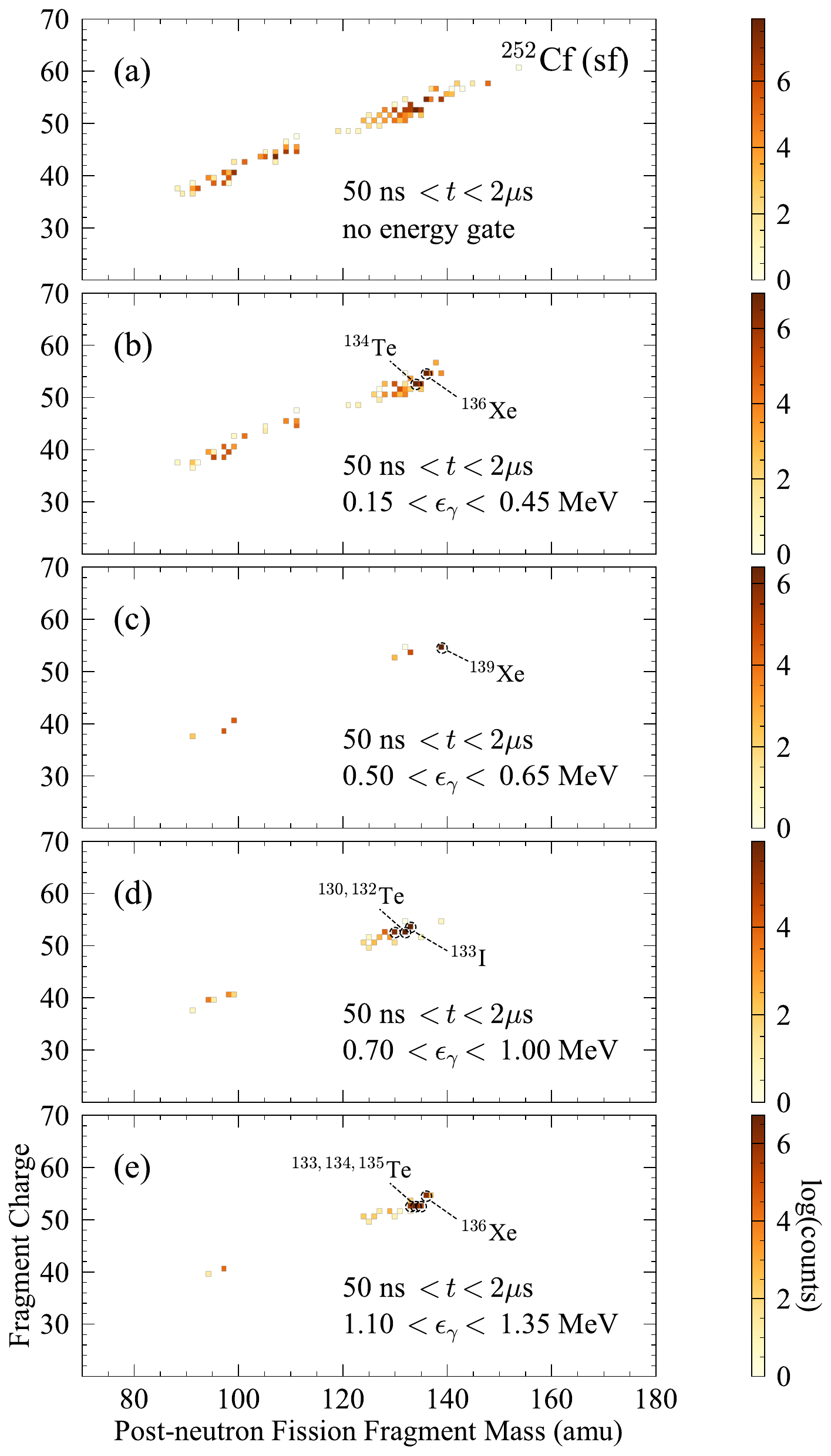}
    \caption{(a) The logarithm of the number of counts of fission fragments (in a \CGMF\ run of 100,000 fission events) that contribute predominantly to emitting late prompt \grays\ in the $^{252}$Cf(sf) reaction within 50 ns to 2 $\mu$s. (b) Same quantity but limited to \grays\ emitted between 0.15 and 0.45 MeV; (c) between 0.5 and 0.65 MeV; (d) between 0.7 and 1.0 MeV; and (e) between 1.10 and 1.35 MeV.}
    \label{fig:chart_gates_Cf252sf}
\end{figure}

Three significant discrepancies between the \CGMF-propagated calculations and the DANCE measurements are seen at the lowest energies below 0.3 MeV, near 0.55 MeV and near 0.75 MeV \gray\ energy. Such discrepancies can be analyzed by studying the \CGMF\ Monte Carlo history files, filtering the \gray\ emission events using the relevant time and energy gates. Figure~\ref{fig:chart_gates_Cf252sf} shows the intensity of the contribution of each fission fragment produced in the $^{252}$Cf(sf) reaction that emits \grays, within 50 ns and 2 $\mu$s following fission, (a) for all energies, (b-e) for different energy windows, as indicated on the figure. 

In the 0.5 to 0.65 MeV energy gate (Fig.~\ref{fig:chart_gates_Cf252sf}c), only one post-neutron fission fragment, $^{139}$Xe, strongly dominates the signal. In the RIPL file of discrete levels used by default in \CGMF, a number of levels with spurious half-lives of 10 ns each were present\footnote{These erroneous half-lives were corrected in the latest RIPL discrete level files after we communicated our findings to the IAEA.}, causing the bulge at 0.55 MeV observed in Fig.~\ref{fig:pfgs_detresp}. 


\begin{figure}
    \centering
    \includegraphics[width=\columnwidth]{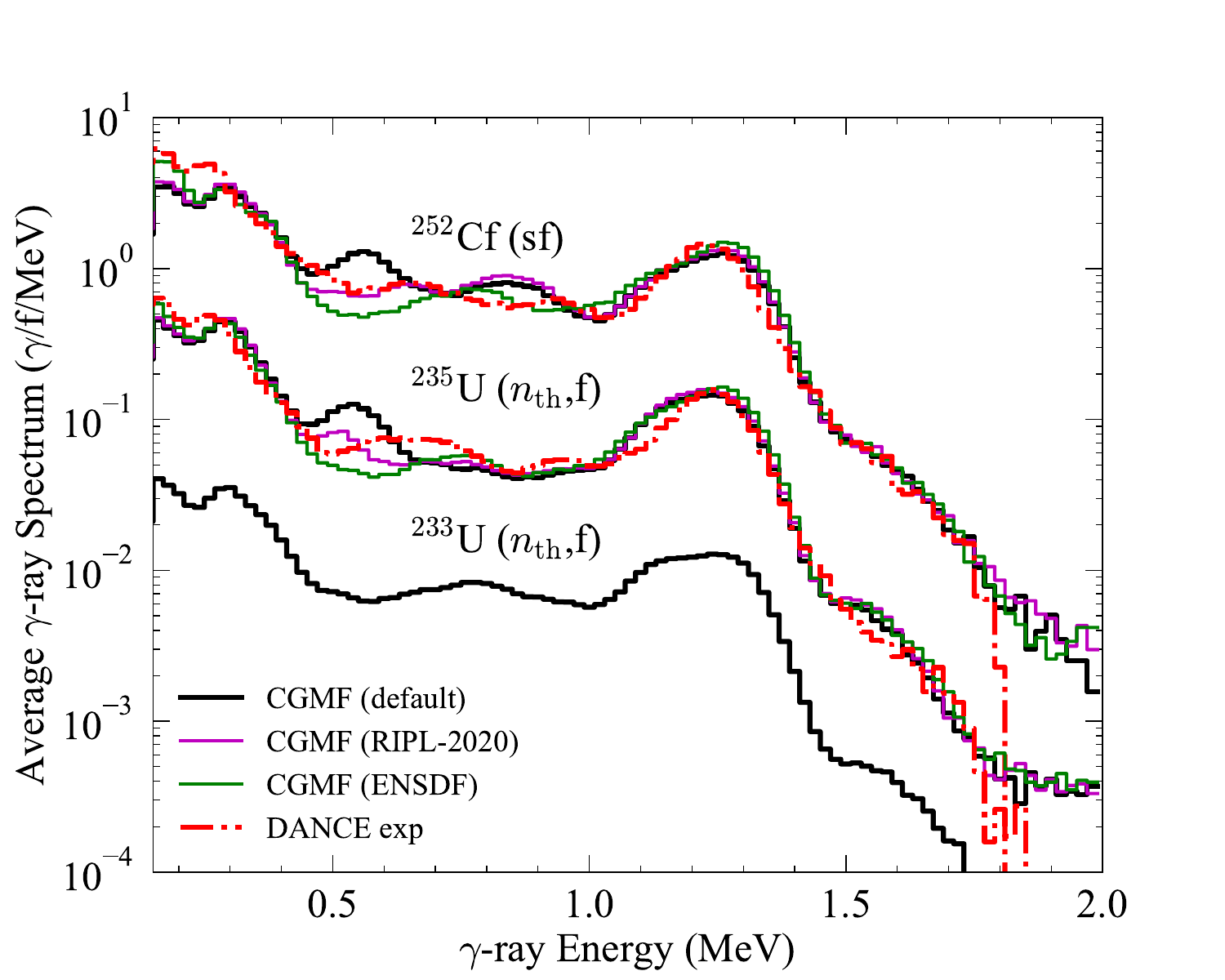}
    \caption{The average \gray\ spectra for the LPFGs in the $^{252}$Cf(sf) and neutron-induced fission reactions on $^{233,235}$U, processed through \GEANT\ and compared to DANCE experimental data (red). Three different level structure files were used in the \CGMF\ calculations: default RIPL-3 from 2015 (black), RIPL-2020 (magenta) and ENSDF (green). The spectra for $^{235}$U and $^{233}$U have been scaled down by factors of 10 and 100, respectively, for clarity.}
    \label{fig:pfgs_detresp_all}
\end{figure}

Any correction in the nuclear structure file used by \CGMF\ should also have an impact on calculations performed for the other fission reactions considered in this work, although those have to be weighted by the initial scission fragment yields $Y(A,Z)$ and neutron emission probabilities P($\nu$). Figure~\ref{fig:pfgs_detresp_all} shows the calculated results (\CGMF\ processed through \GEANT) for all fission reactions studied in comparison with DANCE experimental data. Two different discrete level files were used in the calculations: default RIPL (black) and ENSDF (red). 

Figure~\ref{fig:pfgs_detresp} exhibits another prominent structure near 1.3 MeV in the average prompt fission \gray\ spectrum. \CGMF\ calculations predict two humps near 1.15 MeV and 1.3 MeV for all three isotopes as shown in Fig.~\ref{fig:pfgs_50ns-2us}, which get lumped into just one gross structure once processed through the \GEANT\ detector response simulation. The first peak near 1.15 MeV is mostly due to $^{133}$Te, and to a lesser extent $^{97}$Zr. In $^{133}$Te, the 19/2$^-$ isomer at 1.61 MeV excitation energy has a reported half-life of 100~ns, which first decays to a 15/2$^-$ state, followed by the emission of a 1.151 MeV photon. The peak at 1.3 MeV can be traced back to the presence of the 164~ns 6$^+$ isomer at 1.69 MeV in $^{134}$Te. What is observed is actually the final \g\ transition between the first 2$^+$ state to the 0$^+$ ground state with $\epsilon_\gamma$=1.279 MeV. In both cases, the observed \g\ lines are not stemming directly from the isomeric states but from subsequent decays from lower-lying levels.

These results show that the analysis of the population and decay of these high-spin isomeric states could be used to infer isomeric ratio values in fission fragments before they $\beta$-decay, thereby providing constraints on the angular momentum generated at scission in the fragments.

\subsection{Time-dependent \gray\ Multiplicity} \label{sec:Ngt}

\begin{figure}[hb]
    \centering
    \includegraphics[width=\columnwidth]{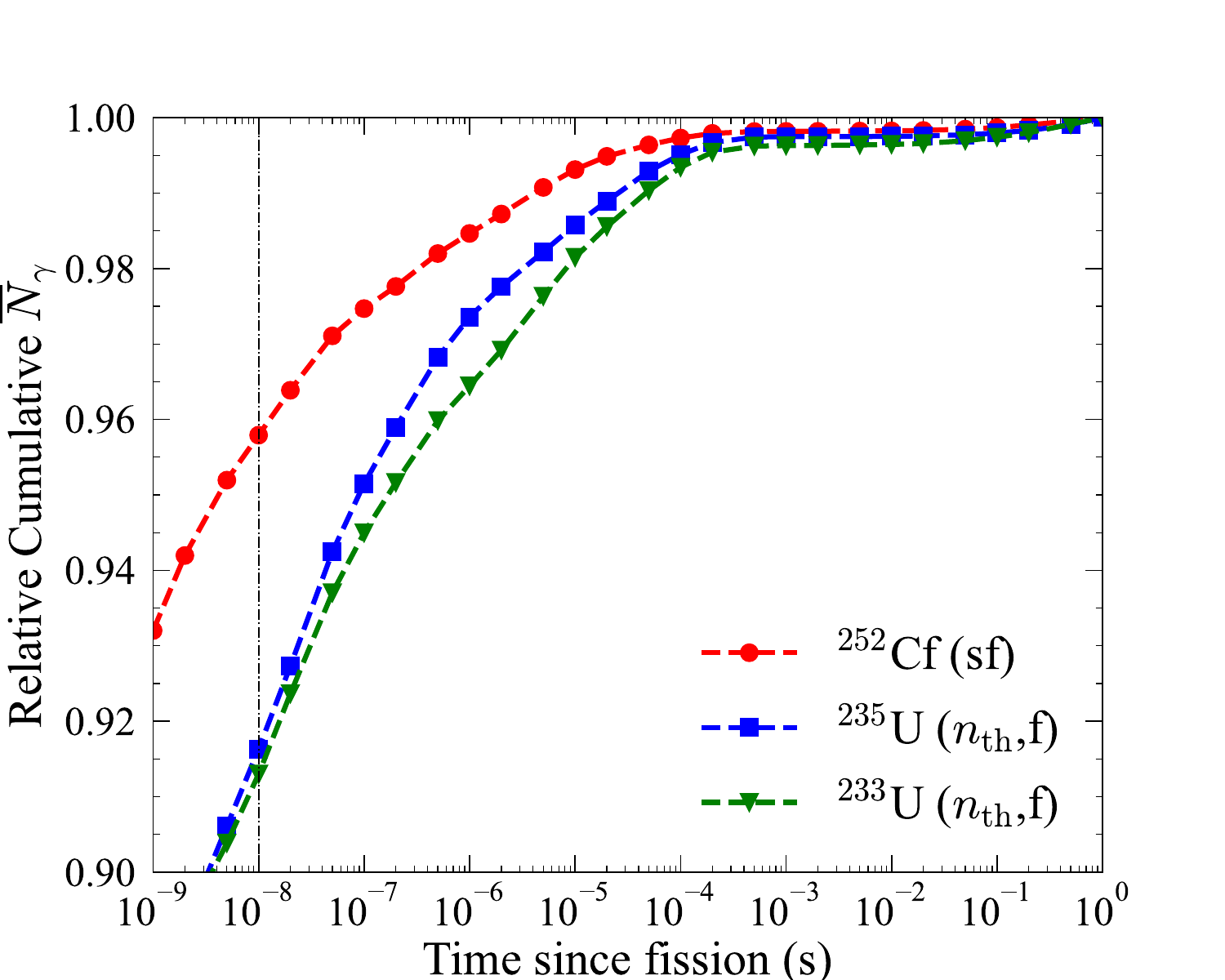}
    \caption{Relative cumulative average \gray\ multiplicity as a function of time since fission for the three fission reactions studied. The dotted-dashed vertical line indicates the 10~ns time, which is commonly used as a time coincidence window for fission event and \gray\ measurements. All curves are normalized to unity at one second after fission.}
    \label{fig:Ng_time}
\end{figure}

Another quantity of interest is the average \gray\ multiplicity, \aveNg, and its evolution over time. Figure~\ref{fig:Ng_time} shows the relative cumulative \aveNg\ as a function of time since fission in the case of $^{252}$Cf(sf) and thermal-neutron-induced fission of $^{233,235}$U. This quantity is normalized to 1.0 at the arbitrary time of one second. As can be seen, about 4\% for $^{252}$Cf(sf) and more than 8\% for $^{233,235}$U ($n_{\rm th}$,f) of the total number of prompt fission \grays\ are predicted to be emitted past the 10~ns mark, which is commonly used in fission experiments to record prompt fission \grays. These numbers should be viewed as upper limits as some of those \grays\ would lie below the energy detector threshold anyway.

\begin{figure}
    \centering
    \includegraphics[width=\columnwidth]{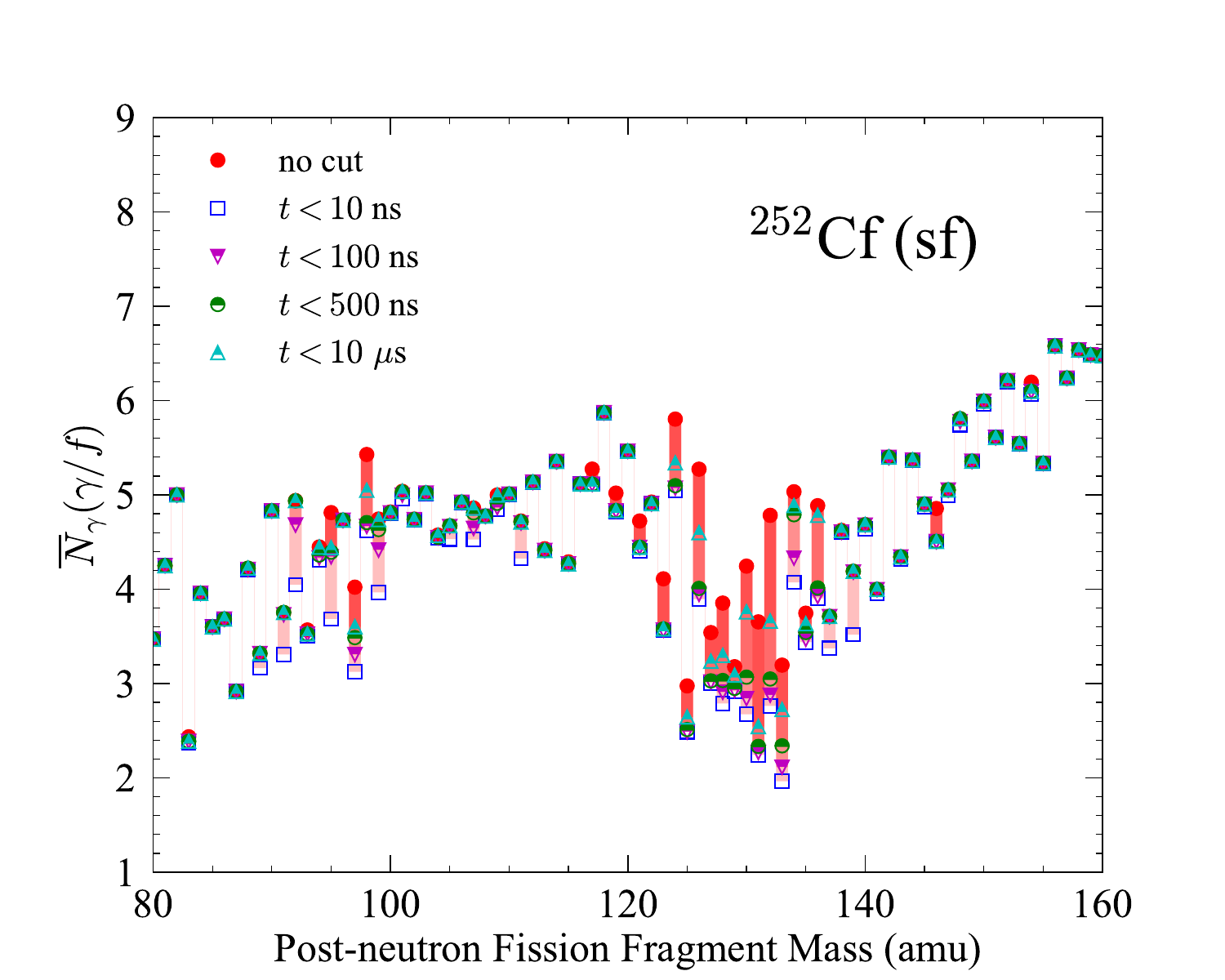}
    \caption{Average \gray\ multiplicity as a function of the fragment mass, for different time cutoffs. The shaded region corresponds to the differences for \aveNg\ calculated with no time cutoff (red disks) and with a time cutoff of 10~ns (blue empty squares).}
    \label{fig:NgAt_Cf252sf}
\end{figure}

While a correction to experimental data reporting \aveNg\ values should be performed, the situation is complicated by the fact that those late prompt \g\ emissions do not occur for all fission fragments equally but only with those that contain the relevant long-lived isomeric states. Figure~\ref{fig:NgAt_Cf252sf} shows \aveNg\ calculated for $^{252}$Cf(sf) as a function of the post-neutron fission fragment mass and for different time cutoffs. 

\begin{figure}
    \centering
    \includegraphics[width=\columnwidth]{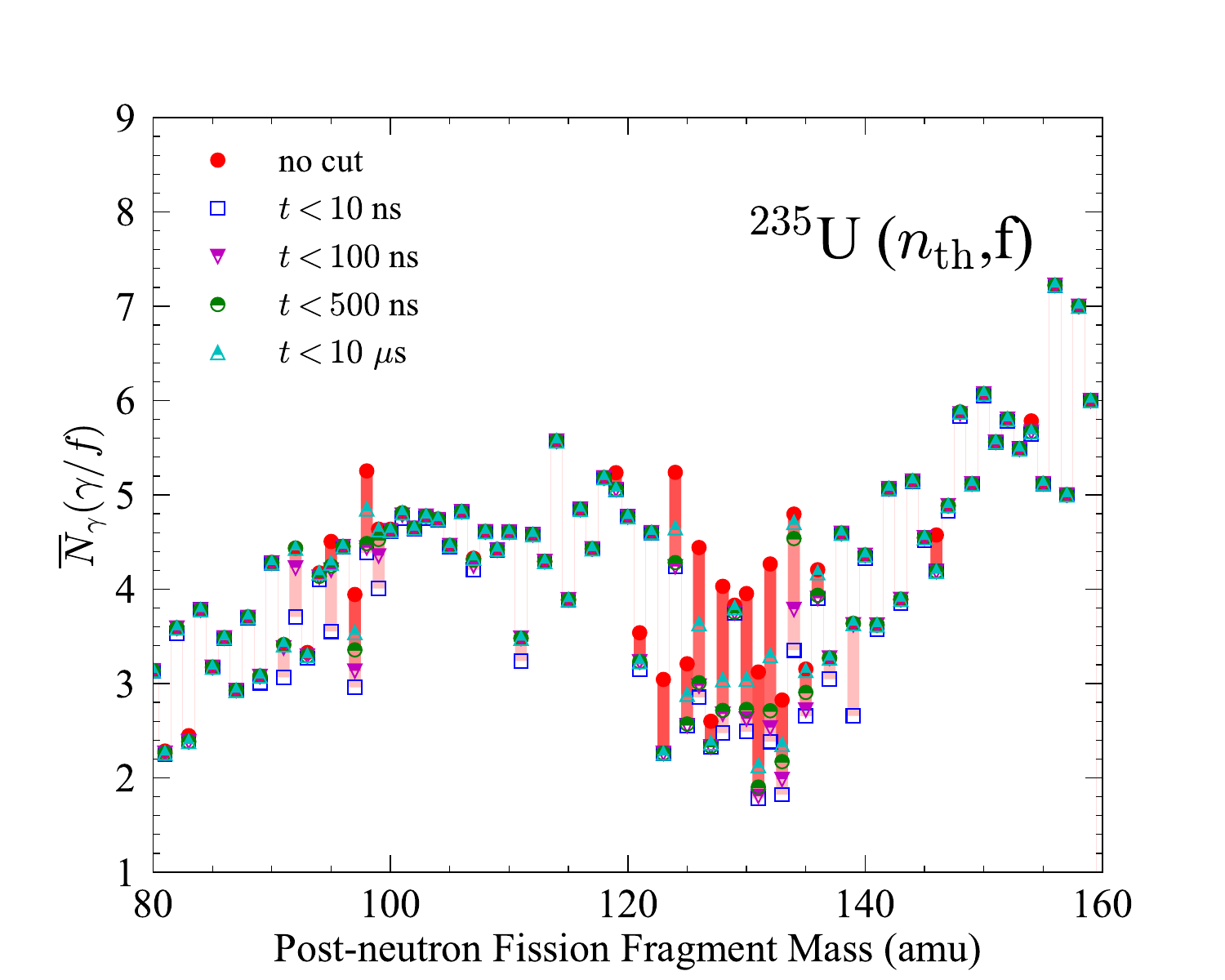}
    \caption{Same as Fig.~\ref{fig:NgAt_Cf252sf} for the thermal-neutron-induced fission of $^{235}$U.}
    \label{fig:NgAt_U235T}
\end{figure}

\begin{figure}
    \centering
    \includegraphics[width=\columnwidth]{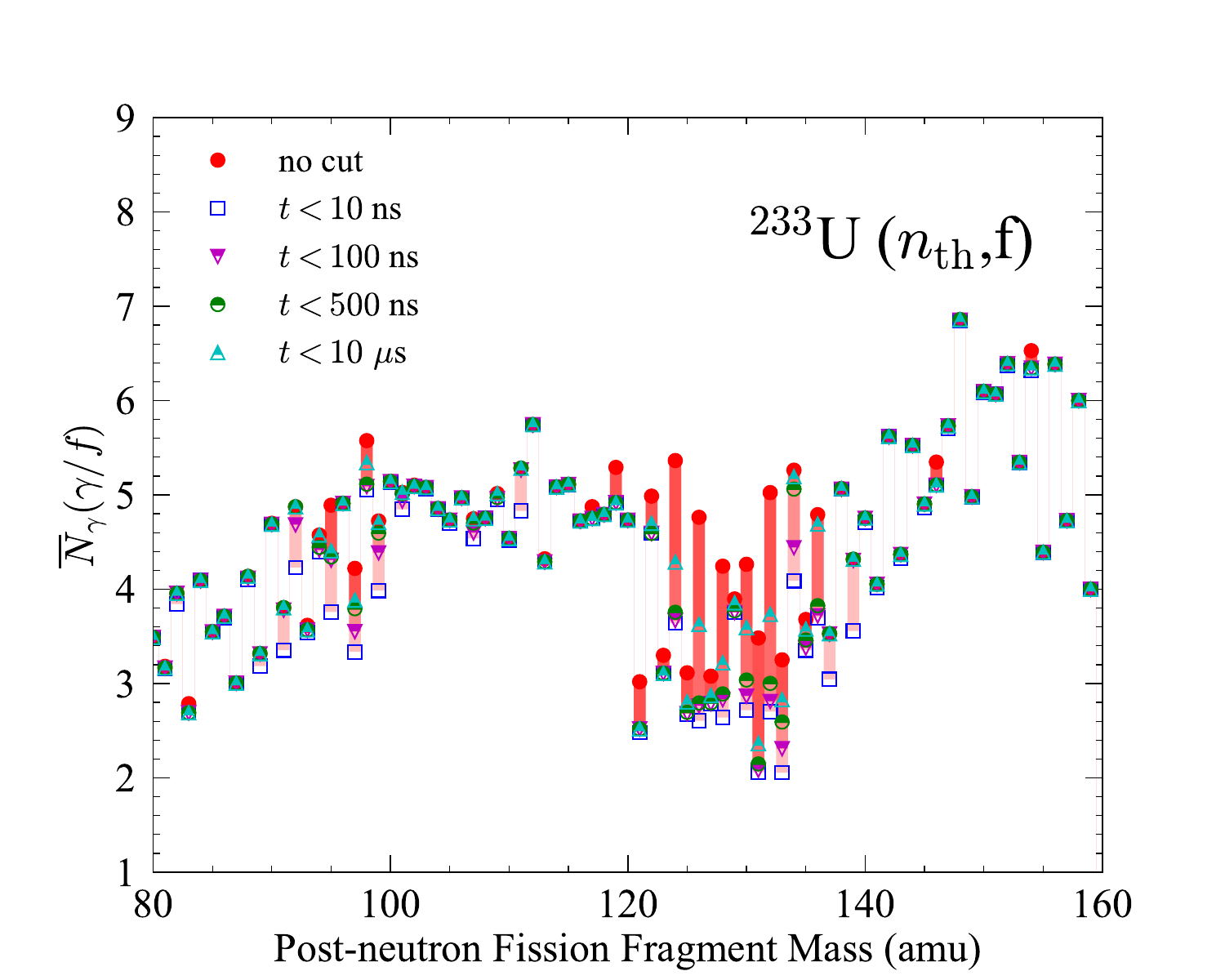}
    \caption{Same as Figs.~\ref{fig:NgAt_Cf252sf} and~\ref{fig:NgAt_U235T} for the thermal-neutron-induced fission of $^{233}$U.}
    \label{fig:NgAt_U233T}
\end{figure}

As can be seen on this figure, most values of \aveNg\ as a function of $A$ of the fission fragment are not sensitive to the choice of the time coincidence window. However, significant differences appear in the [120:140] mass region, and to a lesser extent [95:100], where isomeric states appear and are populated through the fission decay process. These are \CGMF\ predictions and therefore rely on our current knowledge of the nuclear structure of the fission fragments and modeling. The results reported in Figs.~\ref{fig:NgAt_Cf252sf}-\ref{fig:NgAt_U233T} were obtained with the default \CGMF\ input files. Erroneous levels and/or half-lives, as seen in the previous section, could lead to a (slighly) different picture. A measurement of the dependence of \aveNg$(A)$ as a function of the chosen time coincidence window would be valuable.

The same calculation was performed for the thermal-neutron-induced fissions of $^{235}$U and $^{233}$U, and the results are shown in Figs.~\ref{fig:NgAt_U235T} and~\ref{fig:NgAt_U233T} respectively. A similar effect is observed, again concentrated in the [120:140] mass region. The largest differences appear in the case of $^{233}$U. This is consistent with the result shown in Fig.~\ref{fig:Ng_time} where more LPFGs are emitted past 1~$\mu$s in the case of $^{233}$U compared to $^{252}$Cf and $^{235}$U. Those results also confirm our previous discussion~\cite{Stetcu:2021} that showed that the presence of isomeric states in the fission fragments after neutron emission can significantly impact the \g\ multiplicity ``sawtooth" as a function of fragment mass inferred from partial experimental information~\cite{Wilson:2021}.

\subsection{Dependence on the Incident Neutron Energy} \label{sec:En}

The DANCE+NEUANCE experimental data are confined to low incident neutron energies, below a few keV. In fact, we have compared theoretical results for thermal incident neutrons rather than the appropriate experimental neutron spectrum. Within the limitations of our model, in particular the fact that we do not know how to predict the initial scission fragment yields accurately enough to make a distinction between thermal neutrons and a few keV ones, this comparison is valid. However, \CGMF-calculated initial fragment yields are energy-dependent and can impact the predicted characteristics of \gray\ emission as a function of incident neutron energy. 

\begin{figure}
    \centering
    \includegraphics[width=\columnwidth]{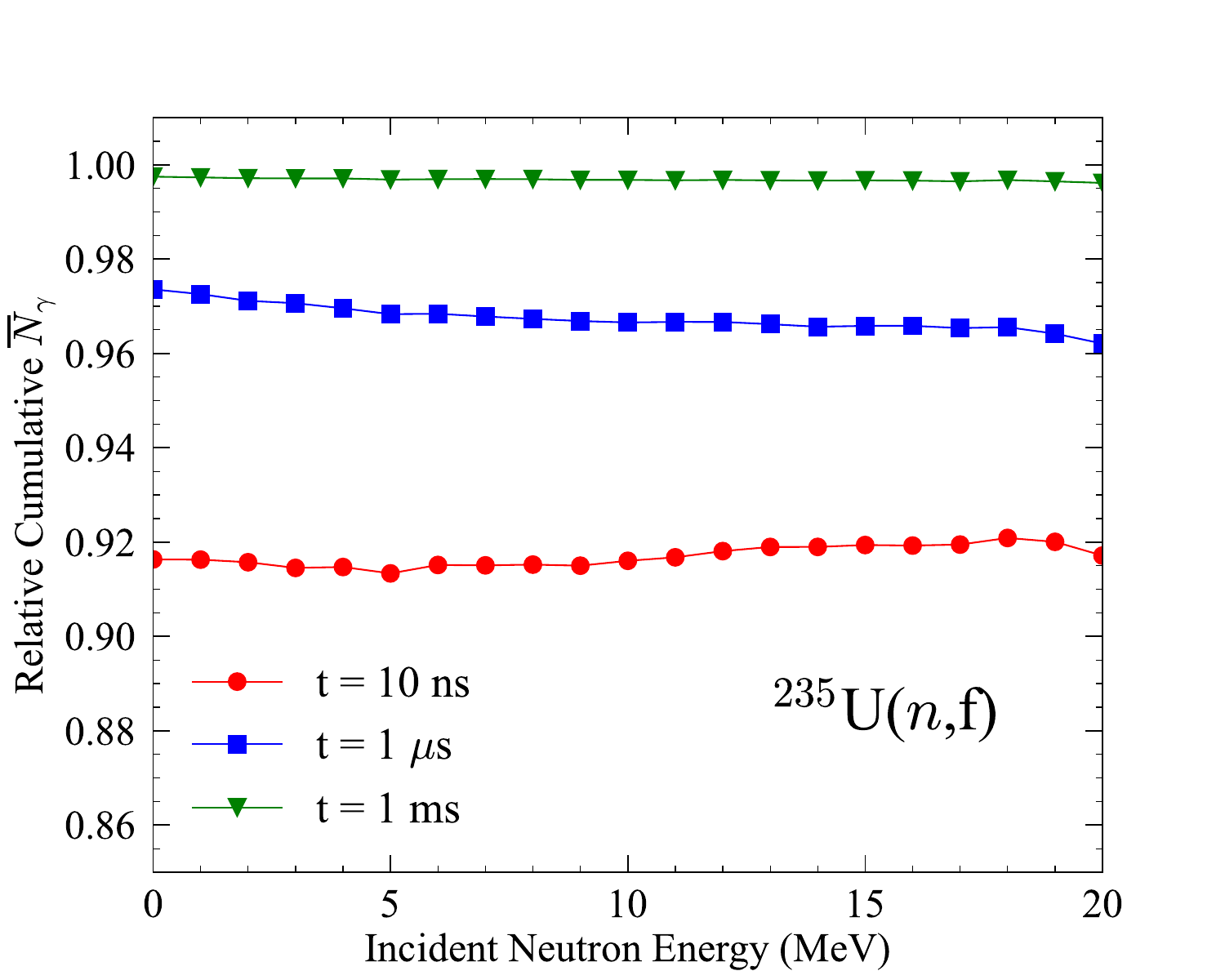}
    \caption{Relative cumulative $\overline{N}_\gamma$ at $t=10$ ns (red disks), 1~$\mu$s (blue squares), and 1 ms (green triangles) for $^{235}$U($n$,f) reactions as a function of incident neutron energy, from thermal to 20 MeV.}
    \label{fig:Ng_t10ns_Einc}
\end{figure}

In Fig.~\ref{fig:Ng_t10ns_Einc}, we show values of the average relative cumulative \g\ multiplicity $\overline{N}_\gamma$ at 10~ns, 1~$\mu$s, and 1~ms after scission as a function of incident neutron energy, for neutron-induced fission reaction on $^{235}$U. The curves for the time coincidence windows of 10~ns and 1~ms remain relatively flat, while the one for 1~$\mu$s decreases slightly with incident neutron energy. As the incident neutron energy increases, the excitation energy in the fragments also increases. Most of this energy is used to emit more neutrons and slightly more energetic neutrons, at least within a particular chance-fission region. Most of the excitation energy left for the \g\ emission remains relatively constant. As seen in Eq.~(\ref{eq:spincutoff}), the spin cut-off parameter $B$ depends on the temperature $T$, but only slightly in our calculations. The relatively constant value for $B$ as a function of temperature or excitation energy explains the results shown in Fig.~\ref{fig:Ng_t10ns_Einc}.

\section{Conclusion} \label{sec:conclusion}

The time evolution of the emission of prompt fission \grays\ has been shown to reveal interesting information about the population or even existence of isomeric states in fission fragments. Despite its limited energy resolution and complex detector response simulation, the DANCE+NEUANCE experimental setup at LANSCE has provided valuable data on the late prompt fission \grays\ which emission is delayed by the half-lives of the isomers that are populated during the fission reaction. We have used the \CGMF\ event-by-event Monte Carlo fission code to reconstruct the time evolution of all \gray\ emissions occurring in the spontaneous fission of $^{252}$Cf and in the thermal-neutron-induced fission on $^{233,235}$U. Deficiencies in the half-lives reported in nuclear structure databases can cause disagreements between experiment and model calculations. Conversely, the identification of such disagreements in the time-gated late-prompt fission \gray\ spectrum can be analyzed to test the quality of nuclear structure databases for neutron-rich nuclei produced in fission reactions. The time evolution of prompt fission \gray\ emission can also help validate the physics models implemented in event generators such as \CGMF. Finally, the intensities of the observed \g\ lines could be used to constrain the angular momentum distributions in the fission fragments in a fashion similar to the isomeric ratios traditionally measured after $\beta$ decay.

\section*{Acknowledgments}

This work was carried out under the auspices of the National Nuclear Security Administration of the U.S. Department of Energy at Los Alamos National Laboratory under Contract No. 89233218CNA000001. 

\bibliography{references}

\end{document}